\documentclass[a4paper,11pt]{article}
\pdfoutput=1 
\usepackage{jheppub} 
\usepackage{epsfig}
\usepackage{caption}
\usepackage{yfonts}
\usepackage{subcaption}
\usepackage{color,soul}
\usepackage{slashed}
\usepackage{float}
\usepackage{amsfonts}
\usepackage{url}
\usepackage{slashed}
\usepackage{accents}
\usepackage{bbm,multicol}
\usepackage{lipsum}
\usepackage{mathtools}
\usepackage{physics}

\usepackage[normalem]{ulem}

\hypersetup{
  bookmarks=true,         % show bookmarks bar?
  unicode=false,          % non-Latin characters in Acrobat?s bookmarks
  pdftoolbar=true,        % show Acrobat?s toolbar?
 pdfmenubar=true,        % show Acrobat?s menu?
 pdffitwindow=true,     % window fit to page when opened
 pdfstartview={FitH},    % fits the width of the page to the window
 pdfsubject={Dark Matter},   % subject of the document
 pdfnewwindow=true,      % links in new window
 pdfcreator={RevTeX},
 colorlinks=true,       % false: boxed links; true: colored links
 linkcolor=red,          % color of internal links
 citecolor=blue,        % color of links to bibliography
 filecolor=black,      % color of file links
 urlcolor=blue,           % color of external links
  }

\title{Revisiting big bang nucleosynthesis with a new particle species : effect of co-annihilation with nucleons}
\author[]{Deep Ghosh}
\affiliation[]{School of Physical Sciences, Indian Association for the Cultivation of Science, 2A and 2B Raja S.C. Mullick Road, Kolkata 700 032}

\emailAdd{matrideb1@gmail.com}
\abstract{In big bang nucleosynthesis (BBN), the light matter abundance is dictated by the neutron-to-proton ($n/p$) ratio which is controlled by the standard weak processes in the early universe. Here, we study the effect of an extra particle species ($\chi$) which \textit{co-annihilates} with neutron (proton), thereby potentially changing the ($n/p$) ratio in addition to the former processes. We find a novel interplay between the co-annihilation and the weak interaction in deciding the ($n/p$) ratio and the yield of $\chi$. Large co-annihilation strength ($G_D$) in comparison to the weak coupling ($G_F$), potentially can alter the number of nucleons in the thermal bath modifying  the ($n/p$) ratio from its standard evolution. We find that the standard BBN prediction is restored for $G_D/G_F \lesssim 10^{-2}$, while the mass of $\chi$ being much smaller than the neutron mass. When the mass of $\chi$ is comparable to the neutron mass, we can allow large $G_D/G_F ~(\gtrsim 10^2)$ values, as the thermal abundance of $\chi$ becomes Boltzmann-suppressed. Therefore, the ($n/p$) ratio is restored to its standard value via dominant weak processes in later epochs. We also discuss the stability of the new particle in an effective theory framework for co-annihilation. Further, the co-annihilation interaction generates elastic scattering of $\chi$ and nucleons at the next-to-leading order. This provides a way to probe the scenario in direct detection experiments, if $\chi$ is accidentally stable over cosmological timescale. }

\begin{document} 
\maketitle
\flushbottom
\section{Introduction and Summary}
\label{sec:sec1}
Big bang nucleosynthesis (BBN) is one of the great achievements of both  the standard model (SM) of particle physics and hot big bang cosmology. The observed primordial abundances of light matter in the universe agree with the theoretical prediction of BBN to a good approximation. One of the important quantities observed in this context is the primordial abundance of helium, i.e. the mass fraction of helium, $Y_P = 0.2449 \pm 0.0040$ \cite{Aver:2015iza}, which is sensitive to the neutron-to-proton ($n/p$) ratio at the initial epoch of BBN. Neutrons and protons are in thermal equilibrium with the cosmic plasma after the QCD phase transition via SM weak processes. Weak processes become inefficient compared to the expansion of the universe at around temperature, $T=1\,\rm MeV$. Subsequently the ($n/p$) ratio is frozen at a value around ($1/6-1/7$)\cite{Kolb:1990vq}. This provides the initial condition for generating the light nuclei in correct abundance at later epochs. It is apparent that any alteration to this ratio due to some new physics phenomena would change the prediction of BBN to a great extent. The addition of a new particle species affects the ($n/p$) ratio broadly in two ways.  

\begin{enumerate}
\item[\bf{I.}] Additional particle species contributing significantly to the energy density of the universe during BBN, changes the expansion rate of the universe which in turn, delays or hastens the freeze-out of neutrons.
\item[\bf{II.}] Chemical processes involving new particle species can alter the ($n/p$) ratio by removing or adding extra nucleons to the thermal bath.  
\end{enumerate}
In scenario {\bf{I}}, the inclusion of an extra relativistic species at the onset of BBN is accounted as the effective number of extra neutrino species, defined as the following.    
\begin{align}
\Delta N_{eff} =\frac{8}{7}\frac{\rho_\chi}{\rho_\gamma}
\label{eq:Neff}
\end{align}      
where, $\rho_\chi$ is the energy density of new particle species, $\chi$  and $\rho_\gamma$ is the photon counterpart. To note, the above definition is applicable untill the freeze-out of electron-positron annihilation, as the relative heating of the photon bath has not yet taken place \cite{Weinberg:2008zzc}. Though the nuclear reactions associated with BBN happen for a range of the photon temperature, for an order-of-magnitude estimate of the temperature and the lower bound on the mass of $\chi$ (discussed in the next section), it is sufficient to take the onset temperature for BBN to be, $T_\gamma \sim 1 {~\rm MeV}$.

This parametrization works well for the species thermally decoupled from the SM plasma before BBN. In ref.\cite{Kolb:1986nf,Boehm:2013jpa} the authors have shown that an additional particle species, strongly coupled either to photons or to neutrinos via elastic scatterings during BBN alters neutrino-to-photon temperature ratio ($T_\nu/T_\gamma$), thereby changing the standard weak interaction rates. As a result the freeze-out time of neutrons gets affected, in turn the final abundances of the light nuclei are altered.

In scenario {\bf{II}}, the standard BBN (SBBN) prediction can potentially be altered due to the infusion of extra nucleons either from the decay \cite{Kawasaki:2017bqm,Holtmann:1998gd,Reno:1987qw,Dimopoulos:1988ue,Kusakabe:2008kf,Yeh:2024ors,Jedamzik:2006xz} or from  the annihilation \cite{Kolb:1981cx,Kawasaki:2015yya,Depta:2019lbe,Hisano:2009rc,Ellis:2011sv} of new particles. In most cases, the decay or annihilation products either inject electromagnetic energy into the cosmic soup via energetic electrons, photons or introduce additional inelastic scatterings between nucleons and nascent antinucleons and other hadrons. The newly formed light nuclei can undergo photo-dissociation due to electromagnetic energy injection, whereas scatterings of hadrons can directly alter the $(n/p)$ ratio, thereby modifying the light nuclei abundances at later epochs \cite{Nollett:2013pwa,Kawasaki:1994sc,Protheroe:1994dt,Kusakabe:2014ola,AlbornozVasquez:2012emy}. Besides, the co-annihilation \cite{Griest:1990kh} of dark sector (with more than one species) particles can give rise to similar effects in light nuclei abundances \cite{Jittoh:2008eq,Jittoh:2010wh,DAgnolo:2018wcn,Ellis:2011sv}.

In this article we have introduced a new particle species $\chi$, which co-annihilates with nucleons. In the co-annihilation process detailed in Sec.\ref{sec:sec2}, a neutron (proton) and a $\chi$ ($\bar{\chi}$) particle are removed from the thermal bath. Therefore, the ($n/p$) ratio and the yield of $\chi$ ($\bar{\chi}$) can simultaneously be affected by the co-annihilation at the relevant epoch. The freeze-out value of the ($n/p$) ratio and the abundance of $\chi$ are decided by the relative strength of the standard weak processes and the newly added co-annihilation, which we have discussed in great details in Sec.\ref{sec:sec3}. In addition, we also discuss the stability of $\chi$ in the context of the effective theory for co-annihilation. In sec.\ref{sec:detection} we discuss the elastic scattering of nucleon (electron) and $\chi$ generated radiatively (detailed in Appendix \ref{app}) from the co-annihilation interaction, leading to a possible probe of accidentally stable $\chi$ in direct detection experiments.

We now summarize our findings regarding the modification of the SBBN scenario due to newly added co-annihilation process. The effect of co-annihilation depends on the initial relative abundances of the neutron (proton) and $\chi$ ($\bar{\chi}$) in the cosmic soup. The number density of neutron (proton) is decided by the observed baryon asymmetry, whereas the number density of $\chi$ is assumed to be thermal. Hence, the number density is decided by the mass of $\chi$ ($m_\chi$) and the temperature of the cosmic soup. For $m_\chi < \mathcal{O}(m_n)$, $m_n$ being the neutron mass, $\chi$ freezes out relativistically, therefore the ambient number density becomes too large compared to the number density of nucleons. Consequently, the co-annihilation processes ($n+\chi \rightarrow p+e^- $ and $p+\bar{\chi}\rightarrow n+e^+$) keep the ($n/p$) ratio fixed at its relativistic equilibrium value, i.e. $R_0 =1$ even at late times altering its SBBN value. This puts a constraint on the co-annihilation strength, i.e. $G_D/G_F \lesssim 10^{-2}$, $G_F$ being the weak interaction. 

The constraint on the $G_D/G_F$ is significantly relaxed for $m_\chi \sim \mathcal{O}(m_n)$, unlike the previous case. This is due to an interesting interplay between the weak interaction and the co-annihilation in deciding the evolution of the ($n/p$) ratio and the yield of $\chi$ ($\bar{\chi}$) over different epochs. Initially ($T>>1\,\rm MeV$) there  is a large number of $\chi$ ($\bar{\chi}$) and for large $G_D/G_F$ ($\gtrsim 10^2$) values the co-annihilation dominates over weak processes. However, the combined co-annihilation reactions of $\chi$ and $\bar{\chi}$ particles keep the ($n/p$) ratio at its SBBN value, $R_0 =1$. Soon for such a large co-annihilation strength the number density of new particle becomes Boltzmann-suppressed. Consequently, the co-annihilation rate per neutron (proton) becomes sub-dominant and the SBBN prediction of the ($n/p$) ratio is eventually satisfied via weak processes in the later epoch. We notice that for $m_\chi \sim \mathcal{O}(m_n)$, we can allow large values of $G_D/G_F$ without altering SBBN predictions. 

In this co-annihilating scenario, $\chi$ can decay to SM states in absence of any symmetry protection, potentially affecting BBN predictions. In the effective model of co-annihilation discussed in the next section, the tree-level decay of $\chi$ is forbidden on the kinematic ground. It might decay via some loop-induced processes, although the lifetime of $\chi$ is larger than the duration of BBN, preserving the main outcome of the previous analysis. However, if the particle is accidentally stable, we can probe the BBN-allowed parameter space with $m_\chi \sim \mathcal{O}(m_n)$ and $G_D/G_F \gtrsim 10^2$ in future direct detection experiments. 
\section{Co-annihilation with nucleons}
\label{sec:sec2}
In the SBBN, the initial number densities of neutron and proton are controlled by the following electro-weak processes in which scattering processes freeze out at $ T\sim 1 \,\rm MeV$ and the $(n/p)$ ratio becomes fixed. The frozen-out value of the $(n/p)$ ratio changes slightly due to occasional decays of neutron until the helium atom is produced, in which most of the neutrons are trapped.
\begin{center}
\bf{A.}  $n+\nu_e \rightleftharpoons p+e^-$,\hspace{0.5cm}
\bf{B.} $n+e^+\rightleftharpoons p+\bar{\nu_e} $,\hspace{0.5cm}
\bf{C.} $n \rightleftharpoons p+e^-+\bar{\nu_e} $
\end{center}
We alter the standard scenario by incorporating a co-annihilation process, i.e. $n+\chi \rightleftharpoons p+e^-$, where $\chi$ is considered as a charge-neutral Dirac fermion. Such a process can be motivated by the lepto-quark mediator models \cite{Belfatto:2021ats} in solving B-physics anomalies as well as solving the neutron decay anomaly \cite{Strumia:2021ybk}. Moreover, the process considered here can be instrumental for detecting sub-GeV particles (dark matter or exotic neutrinos) on the beta-decaying nuclear targets \cite{Dror:2019dib,Long:2014zva,Cocco:2007za}. Apart from being a fundamental particle, $\chi$ can well be a composite particle like neutron, which have been studied in Refs.\cite{Cline:2021itd,McKeen:2020oyr,Carenza:2022pjd} and references therein. Here, we want to investigate the cosmological implication of the process being agnostic about the exact ultra-violet completion of the low energy effective theory, taken as
\begin{align}
\mathcal{L}\supset \frac{G_D}{\sqrt{2}}\left(\bar{\psi}_p \Gamma^a\psi_n\right)\,\left(\bar{\psi}_{e}  \Gamma_a\psi_{\chi}\right)
\label{eq:model}
\end{align}  
where $\Gamma_{a}$ in general, represents all possible independent combinations of Dirac matrices. As a proto-type example of our scenario, We assume the standard $V-A$ structure for $\Gamma^a$, which is reminiscent of SM weak interaction. $\psi_i$ is the fermion field for the $i^{th}$ particle. We want to study only the effect of the co-annihilation on the SBBN case, therefore we take $m_\chi < m_n+m_p+m_e \sim 1.88 \,\rm GeV$ to prevent the tree-level decay i.e. $\chi \rightarrow \bar{n}+ p + e^{-}$ on the kinematic ground.\footnote{There can be loop-induced decay modes of $\chi$ allowed by Eq.\ref{eq:model}, which have been discussed in Sec.\ref{subsec:sec3.2}.} In addition, to exclude the scenario {\bf{I}} discussed in Sec.\ref{sec:sec1}, $\chi$ should not contribute significantly to the energy density of the universe during BBN. If $\chi$ is a decoupled but internally thermalized relativistic species, the measurement of $ N_{eff}$ \cite{Fields:2019pfx} sets an upper bound on the temperature ($T_\chi$) of $\chi$ using Eq.\ref{eq:Neff} as,
\begin{align}
\left(T_\chi/T_\gamma\right) \lesssim 0.6 \hspace{1cm} \text{at 68\% C.L.}
\end{align}
 If $\chi$ is thermalised with the photon bath during BBN, the lower bound on $m_\chi$ comes to be approximately $1.68 \, \rm MeV$. For more detailed analysis on this frontier see Ref.\cite{An:2022sva}, and references therein. In the subsequent analysis, we take the number density of $\chi$ to be thermal to start with. Thus we get a very general mass window for a stable thermalised extra fermion species as, 
\begin{align}
1.68 \,{\rm MeV } \lesssim  m_\chi \leq 1.88\, \rm {GeV}
\label{eq:massbound}
\end{align} 

Now, the co-annihilation process is active once neutrons and protons are available in the bath after the QCD phase transition, $T \sim 150\,\rm MeV$. To check the thermalization condition for $\chi$ via the co-annihilation with neutrons in the expanding universe, we need $\Gamma (T) > H (T)$. Here, $\Gamma (T)$ is the co-annihilation rate per $\chi$ particle and $H (T)$ is the Hubble constant at temperature $T$. The interaction rate depends on the co-annihilation cross-section and the number density of neutrons at a certain epoch. The number density of neutrons is estimated from the observed baryon asymmetry, i.e. 
\begin{align}
\Delta_B = \frac{n_B-n_{\bar{B}}}{s} = \frac{n_n+n_p}{s}=Y_n+Y_p \simeq 10^{-10}
\end{align}   
Where, $Y_i=n_i/s$, $n_i$ is the number density of $i^{th}$ particle and $s$ is the entropy density of the universe. For $T >> 1\, \rm MeV$ the co-moving number density of neutron can well be taken as $Y_n=Y_p=\Delta_B/2$. The s-wave contribution to the thermally averaged co-annihilation cross-section is given by,
\begin{align}
\expval{\sigma v} \cong \frac{G^2_D}{2\pi}\left(m^2_\chi+2m_\chi m_n\right) 
\label{eq:cross}
\end{align}
in which we have assumed electron to be massless and neutron and proton to have identical masses. Assuming usual radiation-dominated universe for the relevant epoch, we arrive at an approximate condition for chemical equilibrium of $\chi$ in the non-relativistic regime as the following. 
\begin{align}
G_D   \gtrsim 4\times 10^{-4} \,{\rm GeV^{-2}} \left(\frac{1 \,{\rm GeV^2}}{m^2_\chi+2m_\chi m_n}\right)^{1/2}\left(\frac{0.1 {\rm GeV}}{T}\right)^{1/2}\left(\frac{10}{g_*}\right)^{1/2}
\label{eq:therm}
\end{align}
Thus, the thermal number density of $\chi$ is ensured around $T \sim 100 ~{\rm MeV} (\sim m_n/10)$, as shown in Figs.\ref{fig:evol1}, \ref{fig:evol2}. The initial production of $\chi$ should happen anytime before $T \sim 100 ~\rm MeV$ either from the SM thermal bath or via the decay of the inflaton field. In addition, $\chi$ can have a non-zero chemical potential related to the lepton number similar to electrons. Therefore, it can take part in leptogenesis\cite{Fukugita:1986hr} and subsequent baryogenesis via the sphaleron effect \cite{Rubakov:1996vz}, if its production happens at sufficiently high temperature, i.e. $T\gtrsim 100 ~{\rm GeV}$. For simplicity, we assume vanishing chemical potential of $\chi$ throughout the cosmic history.   

To study the neutron freeze-out in this modified scenario we need to write Boltzmann equations for neutrons, protons and $\chi$. Due to high entropy density per baryon the light nuclei are not synthesized immediately after the neutron freeze-out, e.g. the synthesis of $He^4$ takes place around $T=0.1\, \rm MeV$. Hence, for $T\gtrsim 0.1 \,\rm MeV$ the relevant processes are only the inter-conversions between neutrons and protons via three weak processes ({\bf A,B,C}) and newly added co-annihilation. As we are considering a thermal number density of the new particle species with vanishing chemical potential, there are two relevant co-annihilation processes involving $\chi$ and $\bar{\chi}$, i.e.   
\begin{align}
1.~ n+\chi \rightarrow p+e^{-}~, ~~ 2.~ p+\bar{\chi} \rightarrow n+e^{+}
\label{eq:co}
\end{align}
The total co-moving number density of baryons is conserved within the relevant epoch, which implies
\begin{align}
\frac{d Y_p}{d x}+\frac{d Y_n}{d x}=0 \, \hspace{1cm}\text{for $T \gtrsim 0.1 \,\rm MeV$}
\label{eq:baryon}
\end{align}  
where $x=m_n/T$. Now, defining the ($n/p$) ratio as $R=Y_n/Y_p$ and $Y_\chi$ ($Y_{\bar{\chi}}$) as the co-moving number density of $\chi$ ($\bar{\chi}$), we write Boltzmann equations using Eq.\ref{eq:baryon} as the following.
\begin{align}
\frac{d R}{d x}&= \frac{1+R}{H x}\bigg[\lambda (p\rightarrow n) -\lambda (n\rightarrow p)R \bigg]-\frac{(1+R)s}{Hx}\bigg[R \expval{\sigma v}_1\left(Y_\chi  -\frac{R_0}{R}\,Y^0_\chi \right)-\expval{\sigma v}_2\left(Y_{\bar{\chi}}  -\frac{R}{R_0}\,Y^0_\chi \right)\bigg]\nonumber\\
\frac{d Y_\chi}{d x}&=-\frac{s \expval{\sigma v}_1}{Hx}\frac{\Delta_B R}{1+R}\bigg[\left(Y_\chi  -\frac{R_0}{R}\,Y^0_\chi \right)\bigg]\nonumber\\
\frac{d Y_{\bar{\chi}}}{d x}&=-\frac{s \expval{\sigma v}_2}{Hx}\frac{\Delta_B }{1+R}\bigg[\left(Y_{\bar\chi}  -\frac{R}{R_0}\,Y^0_\chi \right)\bigg]
\label{eq:boltz}
\end{align}
where $Y^0_\chi$ is the equilibrium co-moving number density for $\chi$, $R_0=Y^0_n/Y^0_p \simeq e^{-Q/T}$ and $Q=m_n-m_p=1.293\, \rm MeV $. $\lambda (n\rightarrow p)$ is the neutron to proton conversion rate via SM weak processes defined assuming zero chemical potential for neutrinos as \cite{Wagoner:1966pv},
\begin{align}
\lambda (n\rightarrow p)=\frac{1}{\tau}\int^{\infty}_{1}d\epsilon \bigg[\frac{\sqrt{\epsilon ^2-1} ~\epsilon ~ (q+\epsilon )^2}{\left(\exp \left(\frac{m_e\epsilon }{T}\right)+1\right) \left(\exp \left(-\frac{ (q+\epsilon)m_e} {T_\nu}\right)+1\right)}+ \frac{\sqrt{\epsilon ^2-1}~ \epsilon  ~(\epsilon -q)^2}{\left(\exp \left(-\frac{m_e\epsilon }{T}\right)+1\right) \left(\exp \left(\frac{m_e(\epsilon-q)} {T_\nu}\right)+1\right)}\bigg]
\label{eq:weakrate}
\end{align}
where, $\tau$ is the neutron lifetime and $T_\nu$ is the neutrino temperature and $q=(m_n-m_p)/m_e=2.53$. The proton to neutron conversion rate, $\lambda(p\rightarrow n)$ is achieved replacing $q$ by $-q$ in the expression of $\lambda (n\rightarrow p)$. In Eq.\ref{eq:weakrate} $T_\nu$ can be written as a function of photon temperature, $T_\gamma = T$ and $T$ is replaced by the scaled temperature variable $x$. Further, $\expval{\sigma v}_i$ is thermally averaged cross-section of process $i$ defined in Eq.\ref{eq:co}. The s-wave contribution to these thermally averaged cross-section can be read off from Eq.\ref{eq:cross} as,
\begin{align}
\expval{\sigma v}_1 \cong \frac{G^2_D}{2\pi}(m^2_\chi+ 2 m_\chi m_n)~~;~~ \expval{\sigma v}_2 \cong \frac{G^2_D}{2\pi}(m^2_\chi+ 2 m_\chi m_p) 
\label{eq:cross-section}
\end{align} 

It is apparent from above Boltzmann equations that both the ($n/p$) ratio and the number density of $\chi$ are inter-related due to added co-annihilation. In particular, $Y_\chi$ ($Y_{\bar{\chi}}$) follows its equilibrium form ($Y^0_\chi$) as long as $R = R_0$, i.e. the ($n/p$) ratio maintains its equilibrium value.  

\section{Results}
\label{sec:sec3}
\subsection{Study of the evolution equations}
\label{sec:dynamics}
We shall now study the evolution of $R$ and $Y_\chi ~(Y_{\bar{\chi}})$ by solving Eq.\ref{eq:boltz} numerically keeping $G_D$ and $m_\chi$ as free parameters. The evolution equations for our scenario are solved upto $x=5000$, which corresponds to $T\simeq 0.2 \,\rm MeV$. As discussed earlier, around $T=0.1 \,\rm MeV$ other nucleons start building up significantly, prompting to include all nuclear reactions into our Boltzmann equations. The temperature range of the current study is adequate to capture general features of co-annihilation. 

It is apparent that in a co-annihilation-type process, the reaction rate per particle is different for each of the colliding particles. Therefore, the co-annihilation process is very much sensitive to the initial relative abundances of two particle species, which is decided by  $m_\chi$. This is because the number density of $\chi$ is thermal and the neutron number density is set by the baryon asymmetry. To demonstrate the effect of new physics on BBN we take two benchmark points (BPs) for  $m_\chi$, i.e. $m_\chi =50 \,\rm MeV$ and $m_\chi = 1.0\,\rm GeV$. 
\begin{figure}[H]
\includegraphics[scale=0.6]{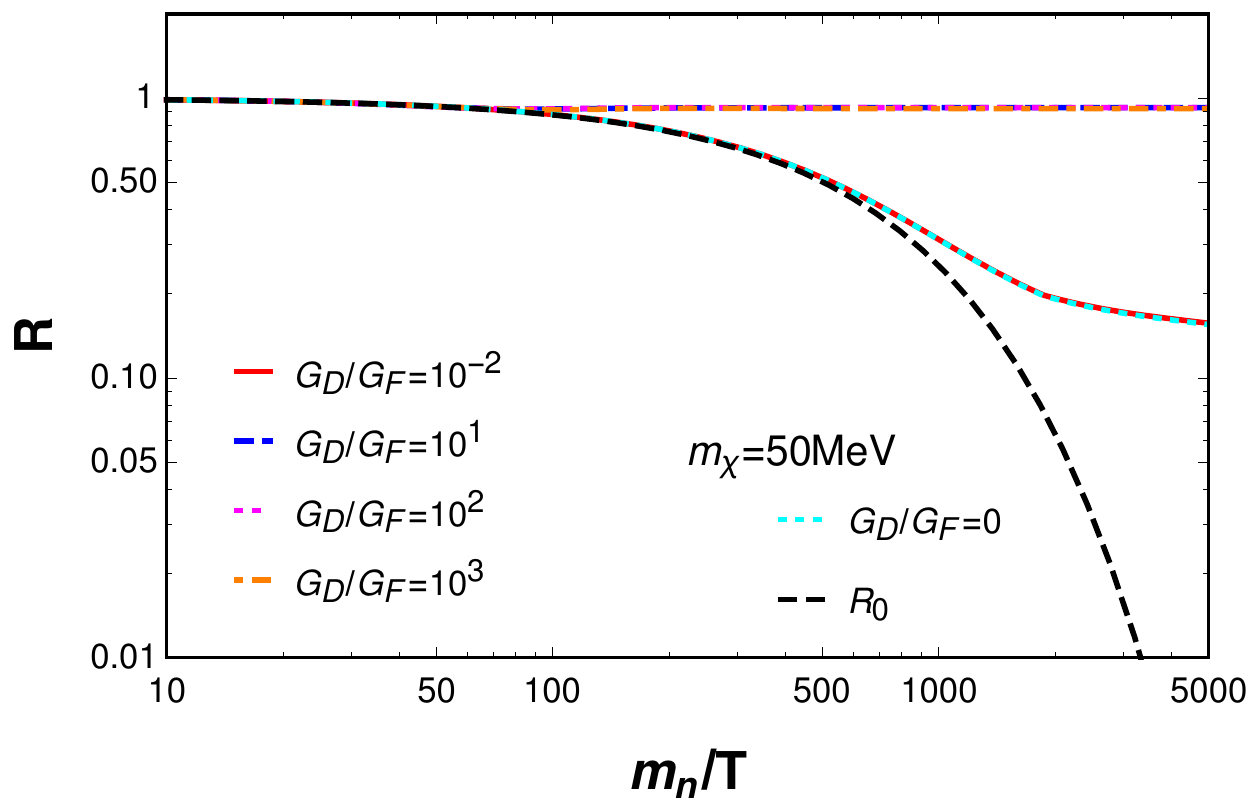}
\includegraphics[scale=.5]{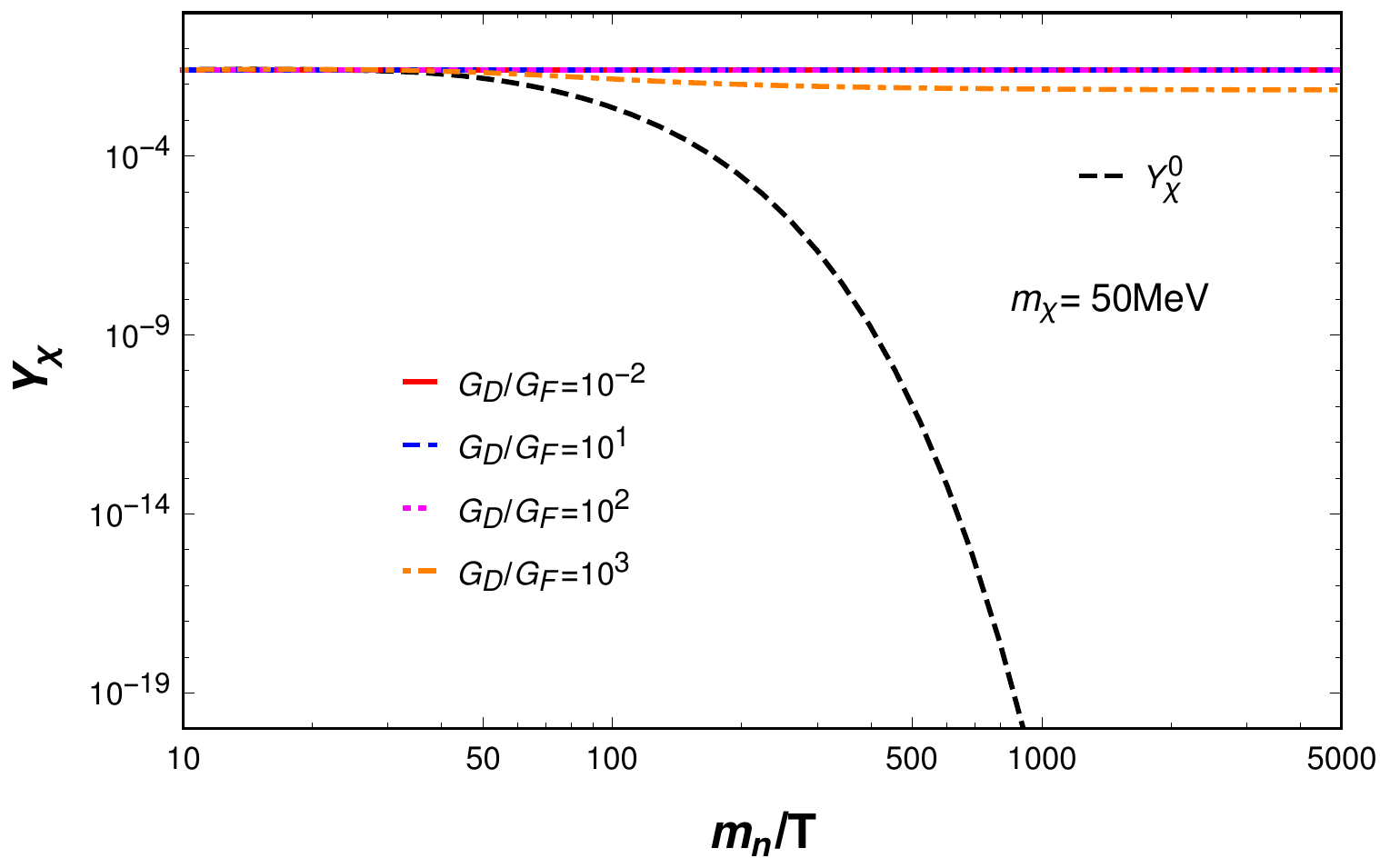}
\caption{Evolution of the ($n/p$) ratio ($R$) (\textit{Left panel}) and the co-moving number density ($Y_\chi$ $(Y_{\bar{\chi}})$) of the additional species (\textit{Right panel}) shown for $m_\chi= 50 \,\rm MeV$, varying the scaled co-annihilation strength, $G_D/G_F$.}
\label{fig:evol1}
\end{figure}
\begin{flushleft}
\textbf{BP - I} : \textbf{Relativistic freeze-out of $\chi$}
\end{flushleft}
In Fig.\ref{fig:evol1} we have shown the evolution of $R$ (left panel) and $Y_\chi$ (right panel) for $m_\chi = 50 \,\rm MeV$ for different values $G_D$, scaled by $G_F=1.166\times 10^{-5} \,\rm GeV^{-2}$. In the right panel, the black dashed line denotes the equilibrium number density ($Y^0_\chi$) of $\chi$ and other lines correspond to the yields with different $G_D/G_F$ values. We find that the additional species, $\chi$ freezes out approximately at its equilibrium value, almost independent of the variation of $G_D/G_F$. For instance, $G_D/G_F$ varying from $10^{-2}$ to $10^{3}$, the freeze-out temperature is approximately same, $T_F \simeq 23\,\rm MeV$ which in terms of the scaled temperature becomes $m_\chi/T_F \simeq 2.2$. This indicates the relativistic freeze-out of $\chi$ as $m_\chi/T_F \lesssim 3$ \cite{Kolb:1990vq}. 

In the left panel, the black dashed line denotes the equilibrium value of the ($n/p$) ratio ($R_0$) and the cyan dashed line corresponds to that of the SBBN scenario, i.e. $G_D/G_F=0$. As long as the co-annihilation keeps $\chi$ in chemical equilibrium, the $(n/p)$ ratio follows the equilibrium value ($R_0$) because there is no extra loss or gain of neutrons via the co-annihilation. After the freeze-out of $\chi$, the ($n/p$) ratio departs from its standard evolution for $G_D/G_F= \{10, 10^2, 10^3\}$, whereas for $G_D/G_F= 10^{-2}$ (red solid line in the left panel) it is unaltered from the SBBN case. This feature can be understood from a comparison between the weak interaction rate ($\Gamma_w$) and the co-annihilation rate ($\Gamma_c$) per neutron as the following.
\begin{align}
\frac{\Gamma_c}{\Gamma_{w}}\simeq \left(\frac{G_D }{G_F}\right)^2 
\label{eq:ratecompare1}
\end{align}       
From Eq.\ref{eq:ratecompare1} we note that for $G_D/G_F= 10^{-2}$ the weak interaction dominates over the co-annihilation throughout the evolution history of neutron and $\chi$. For $G_D/G_F = \{10, 10^2, 10^3\}$  the ($n/p$) ratio remains at its relativistic value due to dominant co-annihilation processes. In particular, the processes, $\chi+n\rightarrow p+e^-$, $\bar{\chi}+p \rightarrow n+e^+$ keep the ($n/p$) ratio around unity even after the chemical equilibrium of the co-annihilation is broken. This is due to a large number of $\chi$ particles present in the cosmic soup compared to  neutrons (protons). In fact, in this case $Y_\chi \simeq 10^{-2}$ and $Y_p=Y_n \simeq 10^{-10}$, which imply that there are $10^8$ number of $\chi$ particles available in $1$ neutron (proton) in the bath. Therefore, the occasional co-annihilation is sufficient to disrupt drastically the neutron freeze-out process as in SBBN. To sum up, for the relativistic freeze-out case, the co-annihilation with larger interaction strength than the weak interaction strength enhances the ($n/p$) ratio by one order of magnitude from the SBBN value at temperature, $T\simeq 10\, \rm MeV$ (see the left panel of Fig.\ref{fig:evol1}). This eventually jeopardizes the predictions of light nuclei abundance, albeit the weak processes are still on.     
\begin{figure}[htb!]
\includegraphics[scale=0.6]{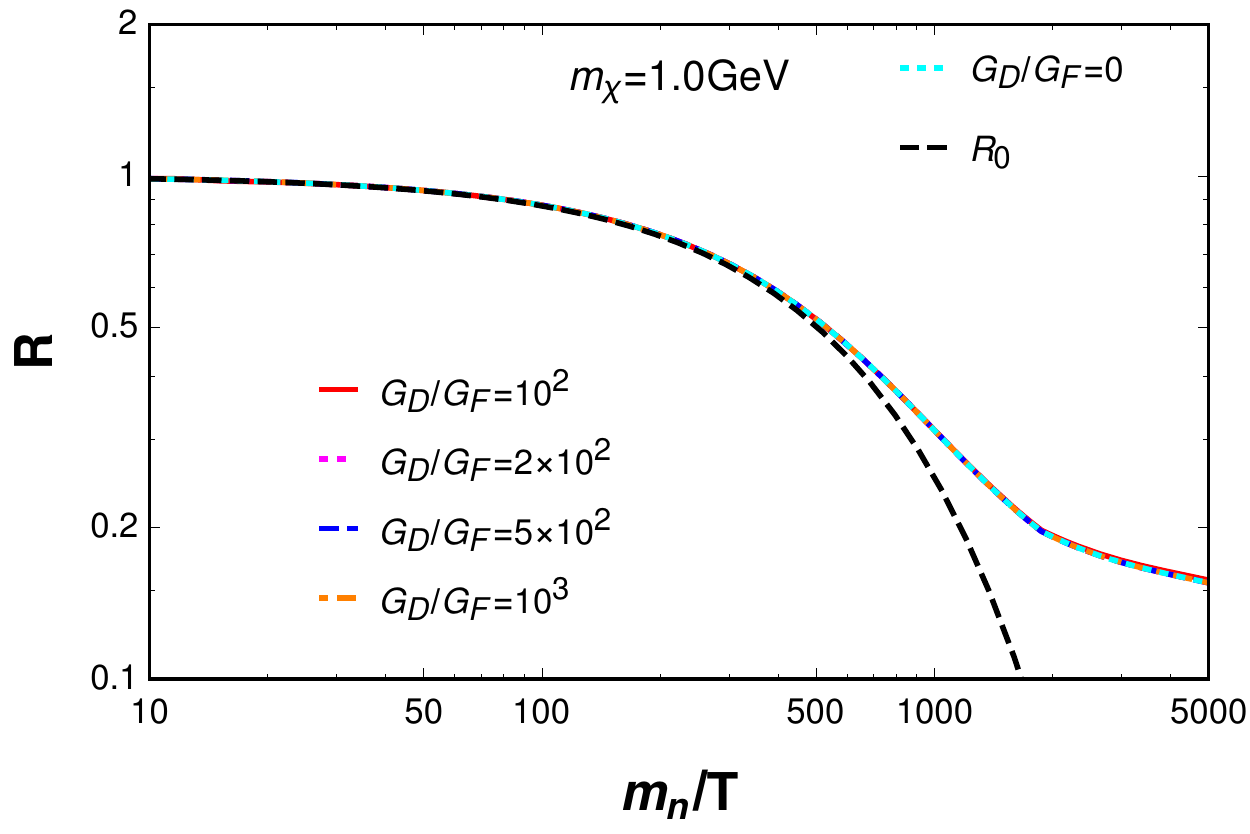}
\includegraphics[scale=0.62]{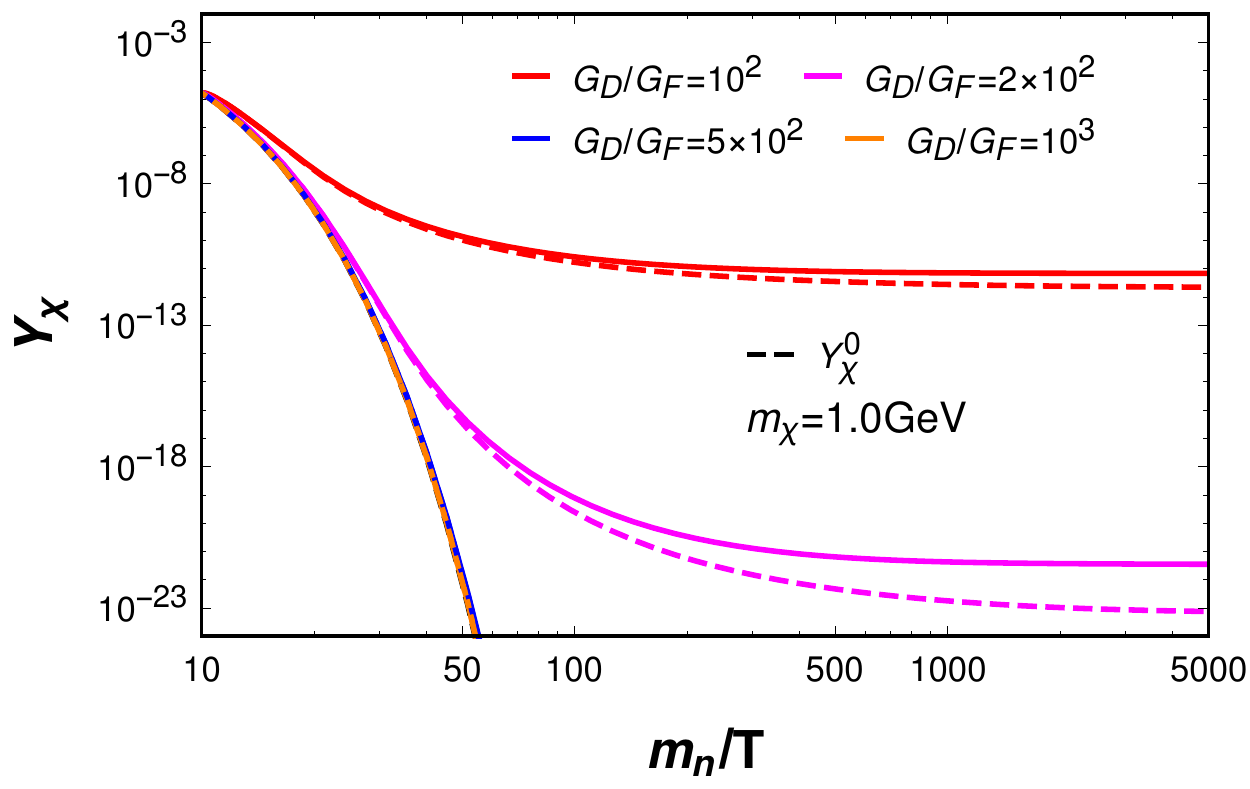}
\caption{Evolution of the ($n/p$) ratio ($R$) (\textit{Left panel}) and the co-moving number density ($Y_\chi$) of the additional species (\textit{Right panel}) have been shown for $m_\chi =1.0 \,\rm GeV$,varying the scaled co-annihilation strength, $G_D/G_F$. The dashed line denotes the co-moving number density of $\bar{\chi}$ in the right panel.}
\label{fig:evol2}
\end{figure}  
\begin{flushleft}
\textbf{BP - II} : \textbf{Non-relativistic freeze-out of $\chi$}
\end{flushleft}
Similar to the previous one, we now show the evolution of $R$ (left panel) and $Y_\chi$ ($Y_{\bar{\chi}}$) (right panel) in Fig.\ref{fig:evol2} for $m_\chi = 1.0\, \rm GeV$ for different values of $G_D/G_F$. Unlike the previous case, the freeze-out of $\chi$ depends on the co-annihilation strength. For example, the freeze-out temperature of $\chi$ for $G_D/G_F = 10^2$ (red solid line) is $T_F \simeq 50 \,\rm MeV$, whereas for $G_D/G_F= 200$ (magenta dashed line for the corresponding yield) the freeze-out happens bit late, i.e. at $T_F \simeq 20 \,\rm MeV$. To note, for $m_\chi = 1\,\rm GeV$ the freeze out of $\chi$ happens in the non-relativistic regime, as $m_\chi/T_F >> 3$ in our example scenarios. In particular, there is a non-trivial relation between freeze-out temperature and the co-annihilation strength, i.e. $T_F \propto G^{-2}_D$ evident from Eq.\ref{eq:therm}. This should be contrasted with the freeze-out temperature in the standard WIMP scenario of which the dependence on the coupling strength is rather weak, i.e. a logarithmic dependence \cite{Rubakov:2017xzr}. The difference manifests from the fact that the freeze-out condition is determined by the baryon asymmetry in our case, whereas in the WIMP scenario it is decided by the thermal densities with vanishing chemical potentials. For $G_D/G_F=\{500, 10^3\}$ (blue dotted and orange dot-dashed line respectively) the co-annihilation keeps $Y_\chi$ in its equilibrium form ($Y^0_\chi$, black dashed line) for the relevant epoch.

In the left panel, We notice that the SBBN scenario for the ($n/p$) ratio ($G_D/G_F =0$ (cyan dashed line)) is reproduced for largish value of $G_D/G_F$ values, i.e.  $G_D/G_F \geq 10^2$. This is a sheer contrast to the relativistic case where largish value of $G_D/G_F$ destroys the SBBN prediction by changing the ($n/p$) ratio significantly. This can be understood by a novel interplay between the weak interaction and the co-annihilation throughout the evolution history of neutron(proton) and $\chi$. To understand the situation, let's look at the relative strengths of the co-annihilation and the weak interaction rate per neutron which is given by the following expression, noticeably different from Eq.\ref{eq:ratecompare1}.
\begin{align}
\frac{\Gamma_c}{\Gamma_{w}}\simeq \left(\frac{G_D }{G_F}\right)^2 \left(\frac{m^2_\chi+2 m_\chi m_n}{T^2}\right) \left(\frac{m_\chi}{T}\right)^{3/2}e^{- m_\chi/T}
\label{eq:ratecompare2}
\end{align}   
At around $T \simeq 90\,\rm MeV$, for $G_D/G_F=10^2$, $\Gamma_c/\Gamma_w \simeq 10^3$, i.e. the co-annihilation rates per neutron (proton) are large by several orders of magnitude from the standard weak processes. Therefore, at this epoch, $\chi+n\rightarrow p+e^-$, $\bar{\chi}+p \rightarrow n+e^+$ decide the ($n/p$) ratio. These processes together do not the alter the the ($n/p$) ratio. As a result the SBBN scenario is restored initially, i.e. $R_0=1$. Very soon, the weak interaction becomes dominant as the number density of $\chi$ is Boltzmann-suppressed. The dominant co-annihilations remove $\chi$ particles substantially from the thermal bath, contrary to the relativistic scenario. Therefore, the available $\chi$ ($\bar{\chi}$) particles per neutron (proton) become scarce to facilitate co-annihilation any further in the later epoch. In particular, at $T \sim 40\,\rm MeV$ for $G_D/G_F=10^2$, $\Gamma_c/\Gamma_w \simeq 10^{-2}$. 

To note, throughout the evolution history, the standard evolution of the ($n/p$) ratio is maintained by two different kinds of reaction. At the early epoch, the co-annihilation processes keep neutrons and protons in thermal equilibrium, subsequently dominant weak processes control the evolution of the same. Hence, in the non-relativistic scenario, largish values of $G_D/G_F$ are allowed from the BBN constraints. In passing, we also note that in the non-relativistic regime, the freeze-out number densities of $\chi$ and $\bar{\chi}$ for $G_D/G_F = \{100, 200\}$ are slightly different. This is due to the fact that the interaction rate of $\chi$ with neutrons is smaller than the interaction rate of $\bar{\chi}$ with protons. In the non-relativistic regime, the number density of neutron is smaller than that of protons due to the mass difference. Now, for $\expval{\sigma v}_1 \simeq \expval{\sigma v}_2$ (see Eq.\ref{eq:cross-section}), the surviving number of $\chi$ particles is larger than $\bar{\chi}$ particles at the freeze-out. Therefore, a small asymmetry is manifested in number densities of $\chi$ and $\bar{\chi}$ particles. 

\begin{figure}[htb!]
\hspace{-0.8cm}
\includegraphics[height=5.5 cm,width=8.5 cm]{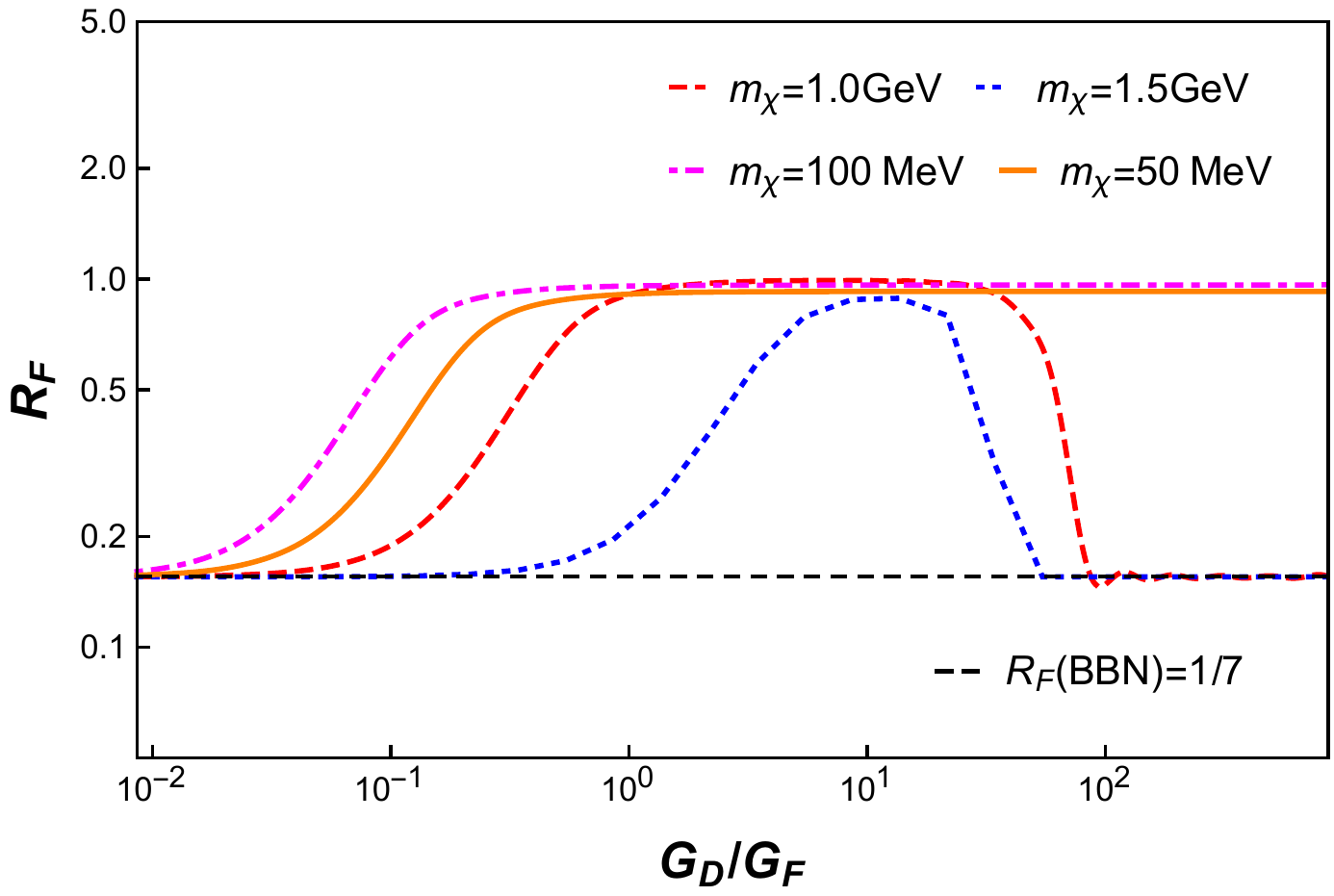}
\includegraphics[scale=0.45]{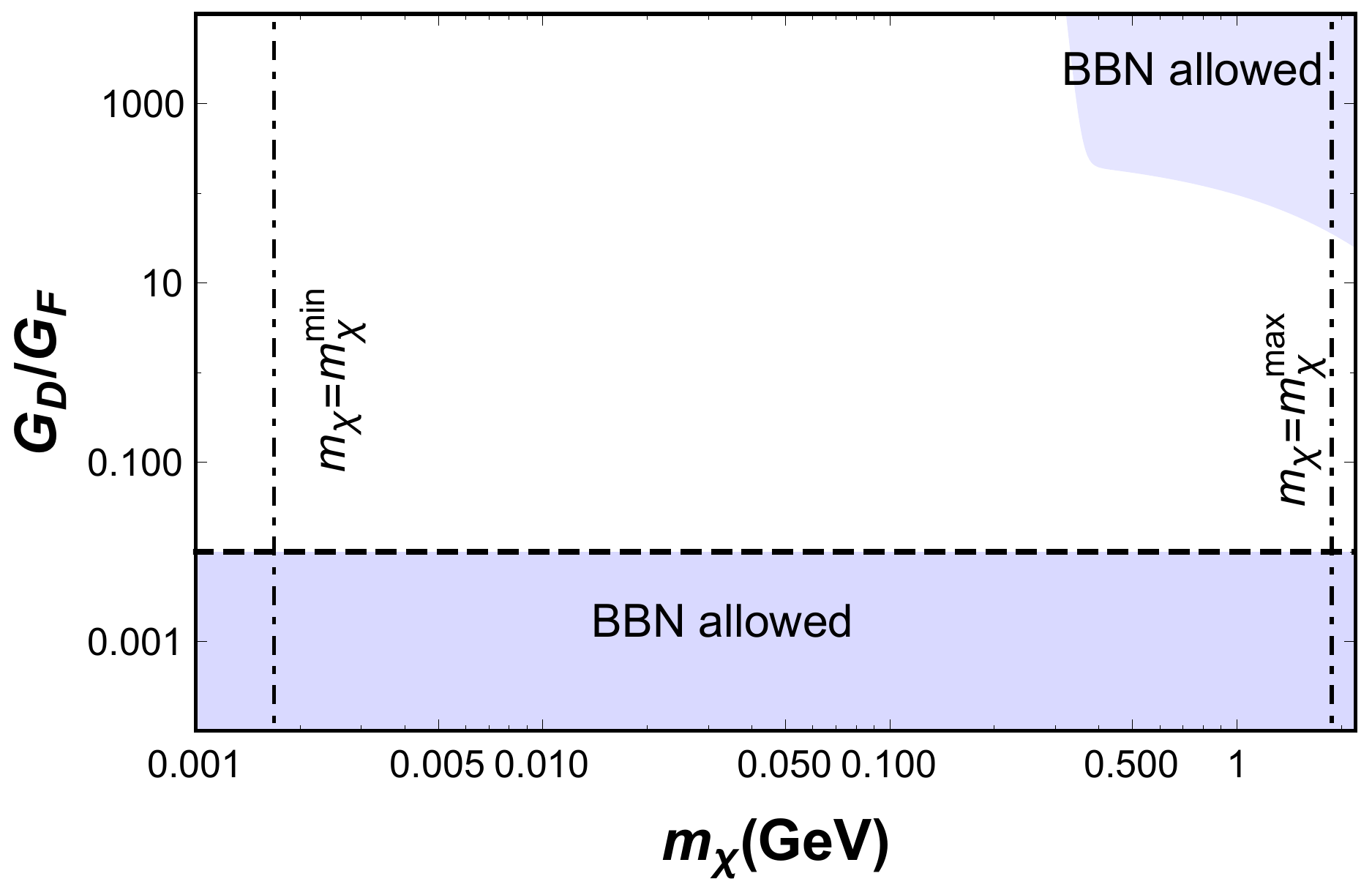}
\caption{Left panel : The freeze-out value of the ($n/p$) ratio, $R_F$ shown as a function of the scaled co-annihilation strength, $G_D/G_F$ for different values of $m_\chi$. Right panel : Allowed parameter space of $m_\chi$ and $G_D/G_F$ shown by the shaded area. }
\label{fig:rfGD}
\end{figure}

We can now summarize our discussion regarding the interplay between the weak interaction and the co-annihilation for both the cases, i.e. the relativistic and the non-relativistic freeze-out of $\chi$ using an instructive diagram shown in the left panel of Fig.\ref{fig:rfGD}. The freeze-out value of the ($n/p$) ratio is denoted by $R_F$, which is calculated at $m_n/T_F = 5000$ varying $G_D/G_F$ continuously over several orders of magnitude for different $m_\chi$ values. In SBBN scenario, $R_F(\rm BBN)\approx 1/7$ \cite{Bernstein:1988ad,Mukhanov:2003xs}, denoted by the black dashed line in the left panel including the effect of the neutron decay. As suggested by the previous two example scenarios, the mass of $\chi$ controls two different features in determining $R_F$.
\begin{enumerate}
\item[\bf{I.}] For $m_\chi << \mathcal{O}(m_n)$, there is a large number of $\chi$ compared to neutrons, which makes weak processes inefficient for  most of the time. Thus $R_F$ deviates from its SBBN value for $G_D/G_F \gtrsim 10^{-2}$. This is illustrated for two masses, i.e. $m_\chi =100\, \rm MeV$ (magenta dot-dashed line) and $m_\chi =50\, \rm MeV$ (orange solid line) in the left panel of Fig.\ref{fig:rfGD}. We note, for $G_D/G_F \approx 10^2$, $R_F \approx 1$, which completely ruins the SBBN predictions.
\item[\bf{II.}] For $m_\chi \sim \mathcal{O}(m_n)$ or larger, $R_F$ remains at its SBBN value for two regions, i.e. for $G_D/G_F \lesssim 10^{-1}$ and for $G_D/G_F \gtrsim 10^{2}$. The SBBN prediction is altered only for the intermediate region, i.e. $10^{-1}\lesssim G_D/G_F \lesssim 10^{2}$. The situation is depicted through two illustrative masses, i.e. $m_\chi =1\,\rm GeV$ (red dashed line) and $m_\chi=1.5 \,\rm GeV$ (blue dotted line). For $G_D/G_F \lesssim 10^{-1}$, the weak interaction dominates over the co-annihilation throughout the evolution history, therefore the SBBN scenario is trivially satisfied. For $G_D/G_F \gtrsim 10^{2}$, the relative abundance of $\chi$ to neutrons (protons) ($Y_\chi/Y_{n/p}$) becomes too small to modify $R_F(\rm BBN)$ via the interplay between the weak and the co-annihilation processes discussed earlier. In the intermediate region, the co-annihilation can not keep $Y_\chi$ in its equilibrium as indicated in Eq.\ref{eq:therm}. Therefore, $Y_\chi/Y_n$ becomes large enough to ruin BBN predictions.      
\end{enumerate} 
To sum up ( as depicted by the shaded area in the right panel of Fig.\ref{fig:rfGD}), when $m_\chi << \mathcal{O}(m_n)$, the BBN prediction allows only small values of $G_D/G_F$. The parameter space for $G_D/G_F$ is enhanced for $m_\chi \sim \mathcal{O}(m_n)$ or larger, as both small and large values of $G_D/G_F$ can reproduce the BBN predictions. We also show the two mass bounds of $\chi$ (as in Eq.\ref{eq:massbound}) by the dot-dashed line in the right panel figure.
\subsection{Probable Loop-induced decays of $\chi$}
\label{subsec:sec3.2}  
\begin{figure}[htb!]
\includegraphics[scale=0.5]{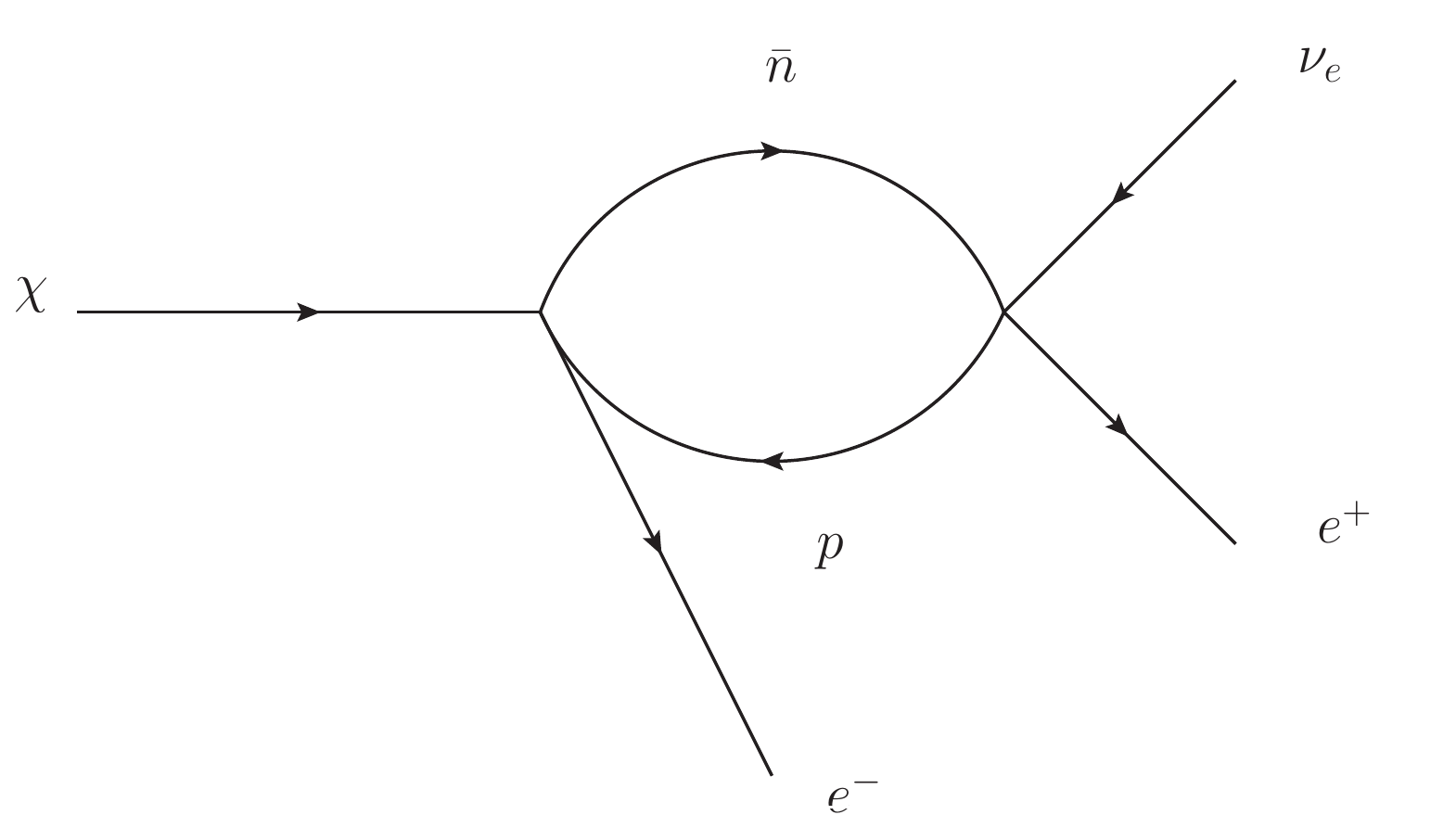}
\includegraphics[scale=0.5]{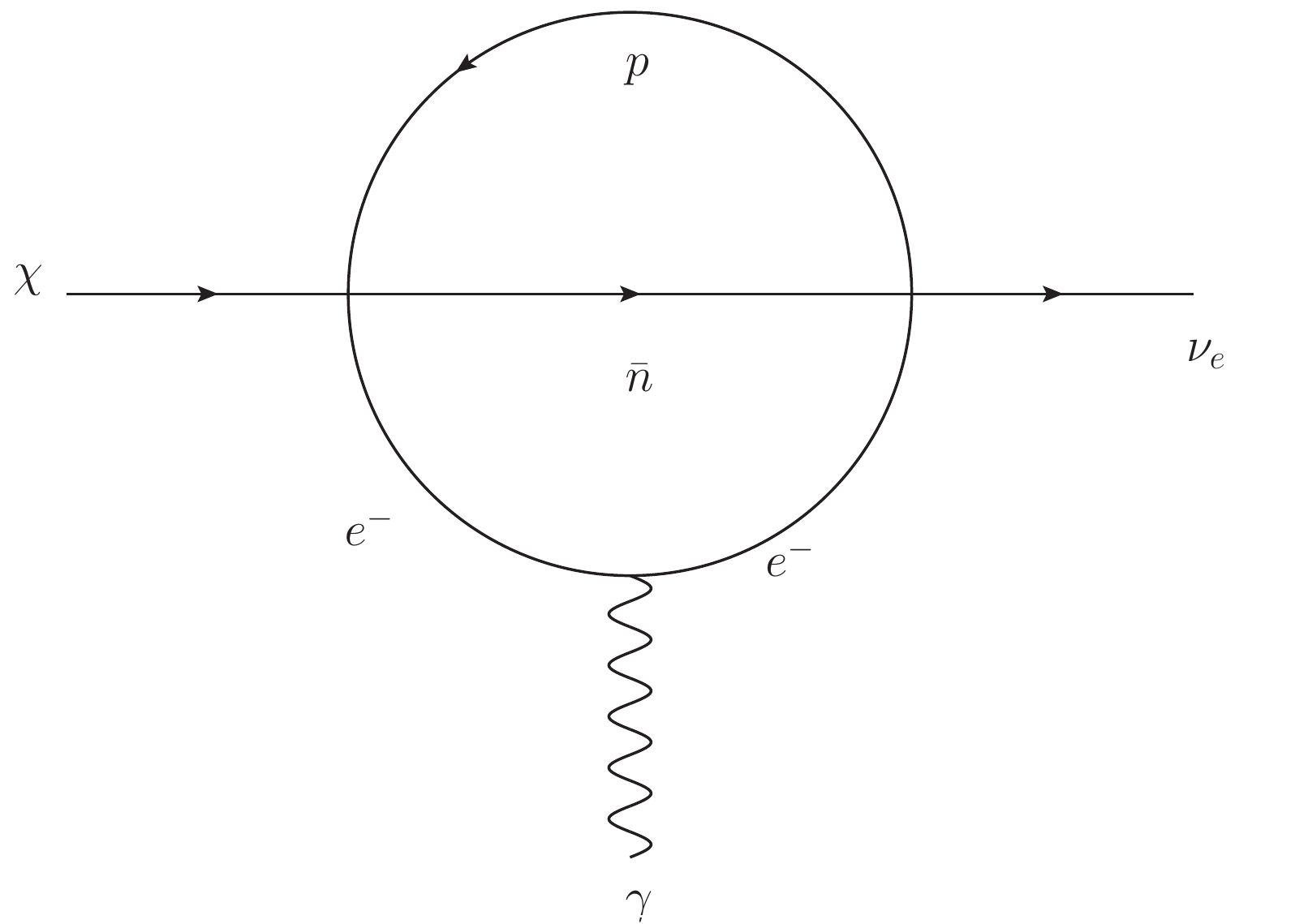}
\caption{Loop-induced decays of $\chi$}
\label{fig:loop1}
\end{figure}

The result shown in the Fig.\ref{fig:rfGD} is modified if $\chi$ decays during the BBN dynamics. Typically, a particle species decaying to electron-positrons, photons or neutrinos modify the hubble expansion rate thus modifying the primordial abundances. In addition, it can change the neutrino-to-photon temperature ratio via energy injection to either of two sectors after the freeze-out of electron-positron pair annihilation. For detailed analysis regarding bounds on the lifetime and the abundance of a decaying particle species during BBN see Refs.\cite{Cyburt:2002uv,Forestell:2018txr,Alves:2023jlo,Hambye:2021moy} and references therein. The co-annihilation effects become prominent when the decay of $\chi$ is sub-dominat during \textit{first three minutes} of the cosmic history. As stated in Sec.\ref{sec:sec2}, the tree-level decay of $\chi$ is avoided on the kinematic ground. However, there can be loop-induced decays as shown in Fig.\ref{fig:loop1}. Within the framework of the effective theory, new interaction terms are generated at quantum level due to these loops, which should be regularized adding appropriate counter terms. Therefore, the decay width of $\chi$ would depend on the residual finite terms of the new interactions generated radiatively after renormalization at a certain order in the perturbation theory.     
Now, we make an order of magnitude estimate of the lifetime of $\chi$ considering only the dominant one-loop decay channel, $\chi (p_1) \rightarrow  e^- (p_2)+e^+ (p_3)+ \nu_e (p_4)$ to check the validity of the parameter space shown in the right panel of Fig.\ref{fig:rfGD}. The contribution from the quadratically divergent one-loop diagram shown in Fig.\ref{fig:loop1} is given by,
\begin{equation}
M_{loop}= (i~ \frac{G_D G_F}{2}) ~\frac{m^2_\chi}{16 \pi^2},
\end{equation}
where we have put $m_\chi$ as the cut-off scale. Now, the decay width of the three body decay is calculated assuming vanishing masses of electron and neutrino as well as having same energies, $E_i=m_\chi/3$.
\begin{align}
\Gamma_\chi \approx \frac{m_\chi}{512 \pi^3} ~|J|^2~|M_{loop}|^2,
\end{align}   
where the contribution from the fermion currents is encoded in $|J|^2 \simeq 0.6~ m^4_\chi$. Then, the lifetime ($\tau_\chi= \Gamma^{-1}_\chi$) of $\chi$ becomes in terms of $m_\chi$ and $(G_D/G_F)$,
\begin{align}
\tau_\chi \approx 10^5 sec \left(\frac{1~ GeV}{m_\chi}\right)^9 \left(\frac{G_F}{G_D}\right)^2 .
\end{align} 
\begin{figure}[htb!]
\includegraphics[scale=0.4]{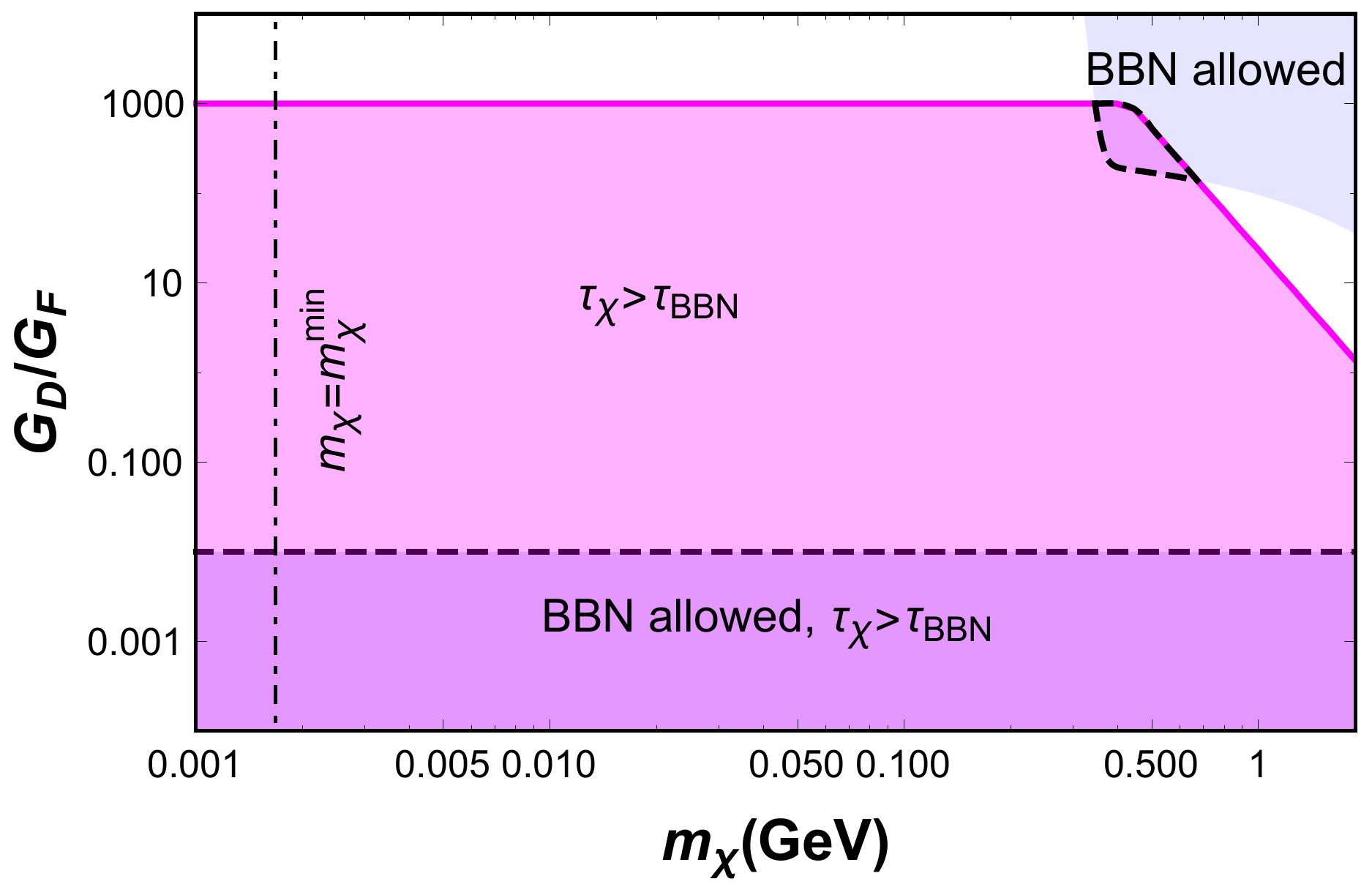}
\includegraphics[scale=0.4]{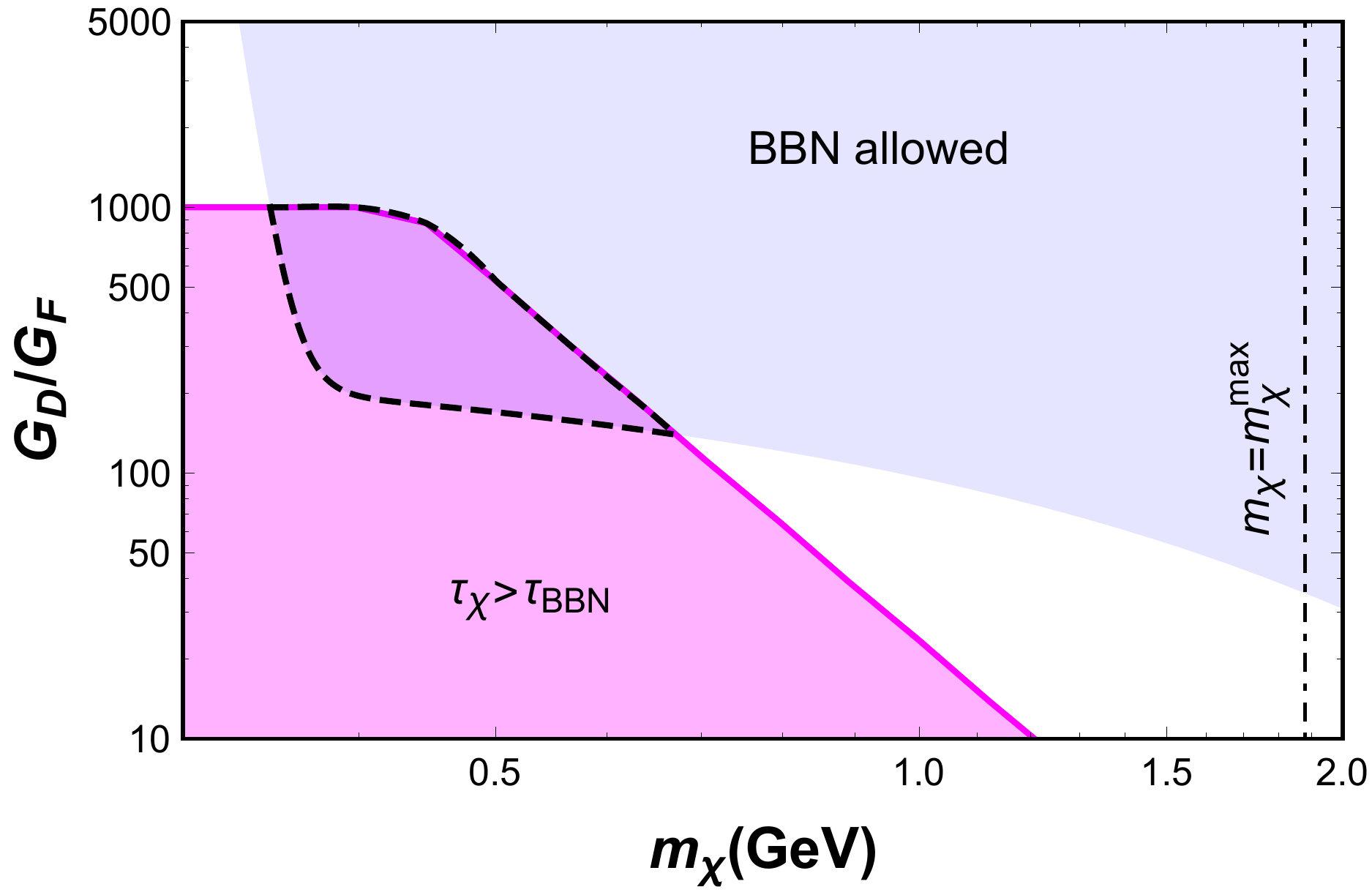}
\caption{Constraints on $m_\chi$ and $G_D/G_F$ considering BBN  results as well as the dominant decay mode of $\chi$. The region below the magenta line indicates the parameter space where the lifetime ($\tau_\chi$) of $\chi$ is greater than the duration of BBN ($\tau_{BBN} \sim 3$ min). In the left panel regions marked by the black dashed line are viable parameter space considering the probable decay of $\chi$. The right panel is the zoomed-in version of the left one, showing larger values for $m_\chi$ and $G_D/G_F$.} 
\label{fig:decaybbn}
\end{figure}

Consequently, our new physics parameters are further constrained as shown in Fig.\ref{fig:decaybbn}, modifying the earlier results as described in Fig.\ref{fig:rfGD} (right panel). In particular, the pink shaded area surrounded by the magenta solid line shows the region in which the lifetime of $\chi$ is greater than the total duration of BBN, i.e. $\tau_{BBN}= 3$ minutes. It is apparent that the for $G_F/G_D \lesssim 10^{-2}$ the allowed parameter space remains unaffected as shown in Fig.\ref{fig:rfGD} (right panel). However, for $G_D/G_F \gtrsim 10^2$, we find the modified mass range of $\chi$ to be $m_\chi \in [0.3,0.7]~\rm GeV$. 

$\chi$ can be cosmologically stable in the regime, $G_D/G_F \lesssim 10^{-2}$ for $m_\chi << \mathcal{O}(m_n)$. For example, $G_D/G_F =10^{-2}$ and $m_\chi = 100 ~\rm MeV$, $\tau_\chi \approx 10^{20}$ sec, which is greater than the current model-independent bound on the lifetime of dark matter (DM), i.e. $5\times 10^{18}$ sec or $160$ Gyr \cite{Audren:2014bca}. However, for such small masses $\chi$ undergoes relativistic freeze-out as explained in \ref{sec:dynamics}, therefore its abundance over-saturates the density of DM in the current universe. For being a valid DM candidate, its density has to be reduced by some other number-changing processes subsequent to the co-annihilation with nucleons. One of the well-studied scenarios in this regard is the cannibalism \cite{Carlson:1992fn,Hochberg:2014dra,Pappadopulo:2016pkp,Ho:2022tbw} of $\chi$ decoupled from the SM thermal bath \cite{Ghosh:2022asg}. The cannibalism can be achieved  incorporating self-interaction of $\chi$ in the effective theory. 

For masses $m_\chi \sim \mathcal{O}(m_n)$, and $G_D/G_F \gtrsim 10^2$, $\chi$ decays after BBN, potentially leaving imprints on the cosmic micro-wave background radiation \cite{Slatyer:2016qyl,Acharya:2019uba} depending on the lifetime of $\chi$.        

In passing, we note that the accidental stability of $\chi$ may be ensured if there is an exact cancellation between diagrams coming from other possible interactions of $\chi$, e.g. similar loops with mesons. The another way to avoid the decay is to think of an extended dark sector \cite{Carenza:2022pjd} where $\chi$ either forms a stable bound state or becomes confined within a DM nuclei or an atom \cite{Cline:2021itd}, i.e. reminiscent of neutron surviving inside the helium nuclei. In a nutshell, we need additional structures in our theory to make $\chi$ absolutely stable. All these require a rather involved study of their own, therefore are left for elsewhere. \\

\begin{figure}[H]
\centering
\includegraphics[scale=0.55]{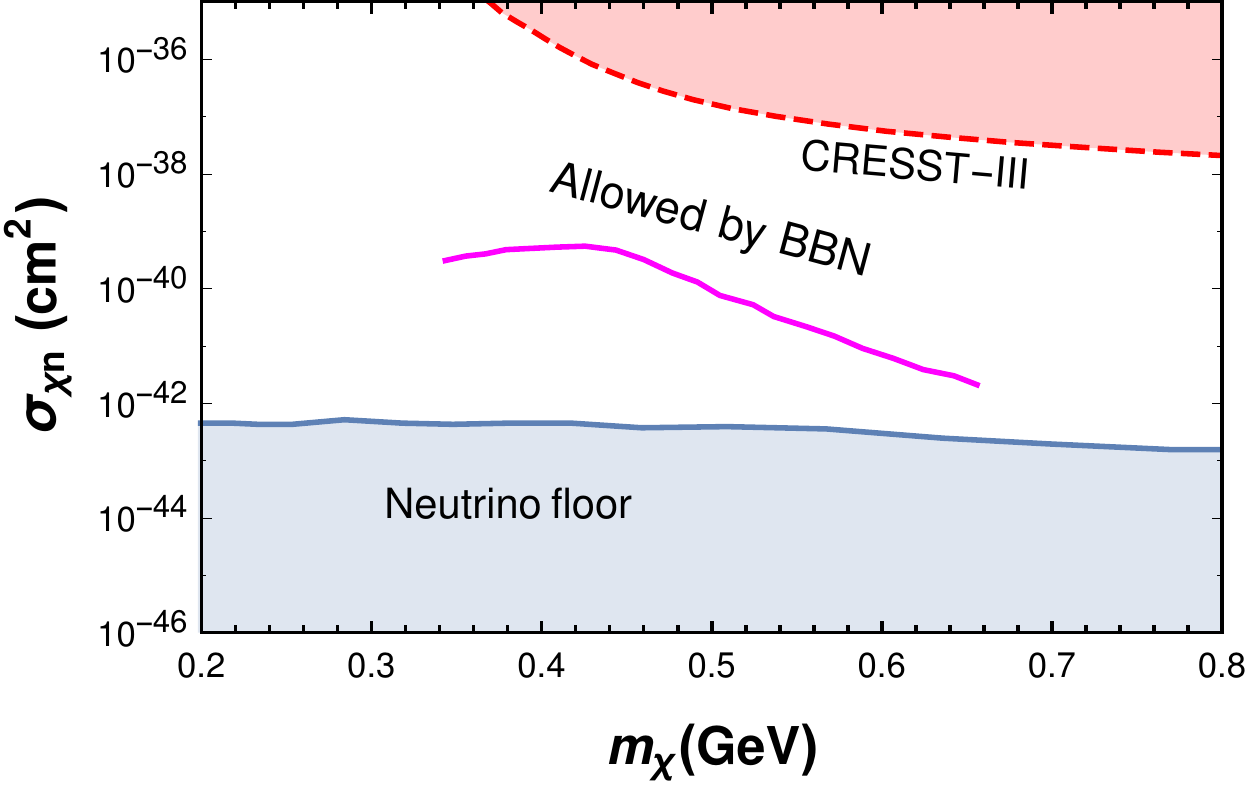}
\includegraphics[scale=0.57]{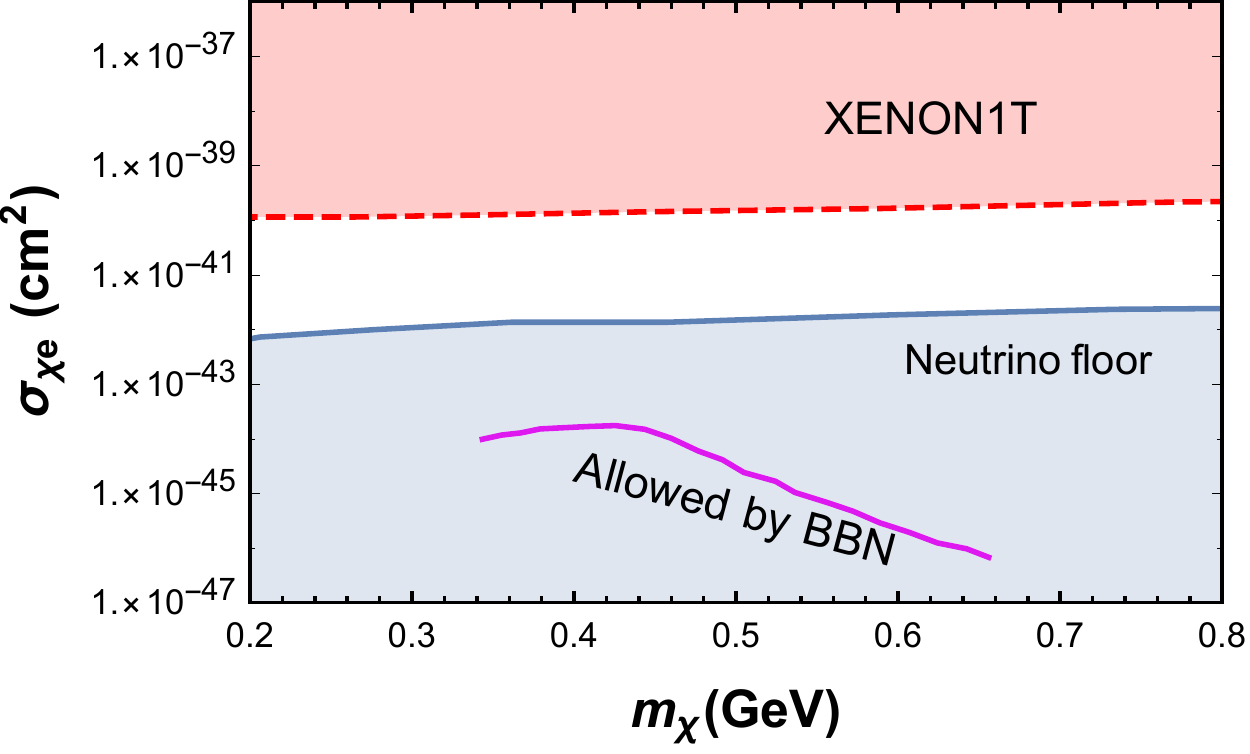}
\caption{DM-nucleon (left panel) and DM-electron (right panel) elastic scattering shown by the magenta solid line in BBN-allowed parameter space depicted in the right panel of Fig.\ref{fig:decaybbn}. The exclusion limit on DM-nucleon scattering given by CRESST-III \cite{CRESST:2017cdd} is shown by the red dashed line. The blue shaded region indicates the neutrino floor where the DM-nucleon scattering becomes indistinguishable with coherent neutrino-nucleus scattering. For DM-electron scattering, similar exclusion limit is shown using XENON1T results \cite{XENON:2019gfn}. The corresponding neutrino floor for liquid xenon is also shown as in Ref.\cite{Carew:2023qrj}.}
\label{fig:direct}
\end{figure}

\section{Elastic scattering of nucleon (electron) and $\chi$}
\label{sec:detection}

The elastic scattering between the nucleon (electron) and $\chi$ is generated at the one-loop level as shown in Fig.\ref{fig:loop}, given the co-annihilation interaction in Eq.\ref{eq:model}. For both $\chi-n$ and $\chi-e^-$ scattering, the scattering amplitudes take the following generic form.
\begin{align}
iM= \frac{i G^2_D}{32\pi^2}\left(A P_\rho P_\sigma+B\frac{\Lambda^2}{4}\eta_{\rho\sigma}\right) \bigg[\bar{u}(p_3)\,\Gamma^\mu \,\gamma^\rho\Gamma^\nu u(p_1)\bigg]\, \bigg[\bar{u}(p_4)\,\Gamma_\mu \,\gamma^\sigma\Gamma_\nu u(p_2)\bigg],
\label{eq:amp}
\end{align}
where $A$, $B$ are the loop factors and $\Lambda$ is the relevant cut-off scale. $\Gamma^\mu$'s are of $V-A$ structure as used before and $P^\mu=(p_1+p_2)^\mu$. To note, loop factors would be different for two scatterings. For details see Appendix \ref{app}. 
We have taken scatterings with neutrons only although similar scattering can be considered with protons as well. Now, if $\chi$ becomes accidentally stable due to some hidden sector dynamics as indicated earlier, these elastic scatterings become instrumental in probing the co-annihilation scenario in terrestrial direct detection experiments. In particular, for $m_\chi > \mathcal{O}(GeV)$, the DM-nucleon scatterings are important \cite{Goodman:1984dc}, whereas the DM-electron scatterings put stringent constraints for sub-GeV DM masses \cite{Essig:2011nj}. In Fig.\ref{fig:direct}, we have shown these scattering cross-sections, allowed from BBN prediction and compared with the present experimental bounds. 

At first, the spin-independent part of DM-nucleon scattering is extracted from the generic expression shown in Eq.\ref{eq:amp}, as the spin-dependent part is always suppressed in presence of the former one. Therefore, the relevant cross-section (detailed in Appendix \ref{app}) is given by    
\begin{align}
\sigma_{\chi n} \approx \frac{ G^4_D}{\pi^5} \bigg(m_\chi m_n (m_\chi+m_n) \bigg)^2 
\end{align}
In our scenario, $\chi$ might account for low mass DM ($m_\chi \lesssim 3 ~\rm GeV$) for which the bound on the spin-independent DM-nucleon cross-section is rather weak for XENON1T-like experiments \cite{XENON:2019gfn}. This type of scenario can be probed at CRESST-III experiments \cite{CRESST:2019jnq} (and via Migdal effect \cite{Bell:2019egg,Dolan:2017xbu}) of which the current bound on the DM-nucleon scattering cross-section is $\sigma_{\chi n} \lesssim 10^{-38} ~\rm cm^2$ for $m_\chi \lesssim 1\,\rm GeV$.

In the left panel of Fig.\ref{fig:direct}, we have shown the detection prospect of BBN-allowed parameter space for $\chi$ particles. There are two regions of interest as in previous section, namely $m_\chi << \mathcal{O}(m_n)$ and $m_\chi \sim \mathcal{O}(m_n)$. For $m_\chi << \mathcal{O}(m_n)$, the BBN prediction allows only small $G_D/G_F  (\lesssim 10^{-2})$ values, thereby $\sigma_{\chi n}$ is well below the neutrino floor (shaded region below the blue solid line). For example, with $m_\chi =0.05 ~\rm GeV $ and $G_D/G_F =10^{-2}$, the cross-section becomes $\sigma_{\chi n} = 5\times 10^{-61}~\rm cm^2$. For $m_\chi \sim \mathcal{O}(m_n)$, the BBN prediction allows for both small and large values of $G_D/G_F$. Therefore, the detection possibility of $\chi$ becomes more optimistic for $m_\chi \in [0.3,0.7]~\rm GeV$ and $G_D/G_F \gtrsim 10^2$. For a benchmark point, $m_\chi = 0.6 ~ \rm GeV$ and $G_D/G_F=250$, we get $\sigma_{\chi n}\approx 9.8\times 10^{-41} ~\rm cm^2 $, which is evidently above the neutrino floor. The magenta solid line in Fig.\ref{fig:direct} depicts the coveted sensitivity of the future experiments to probe the co-annihilation scenario, being consistent with the cosmological observations. The current exclusion limit on the DM-nucleon cross-section coming from the preliminary results of CRESST-III phase-I experiments \cite{CRESST:2017cdd} have been indicated by the red dashed line and the shaded region above the line. 

Now, the DM-electron scattering cross-section can be calculated (detailed in Appendix \ref{app} ) in terms of DM-nucleon scatterings as both type of scatterings are generated radiatively from the same effective co-annihilation operator. The DM-electron scattering cross-section is suppressed compared to the DM-nucleon scattering due to small electron mass, i.e.
\begin{align}
\sigma_{\chi e}\approx \frac{m^2_e}{m^2_\chi m^2_n}  (m_\chi+m_n)^2 \sigma_{\chi n} .
\end{align}
In the right panel of Fig.\ref{fig:direct}, we have shown the BBN-allowed parameter space for DM-electron scattering, contrasted with results from ionization signal of XENON1T experiments \cite{XENON:2019gfn}. Unlike the DM-nucleon scattering, the allowed contour in $\sigma_{\chi e}-m_\chi$ plane is within neutrino floor, making the distinction between the signal and background events difficult. The dominant background comes from coherent neutrino-nucleus scattering, shown as the neutrino floor. We have adopted the neutrino floor line as in Fig.1 (left panel) of Ref.\cite{Carew:2023qrj}. However, the neutrino floor (sometimes called as neutrino `fog'\cite{OHare:2021utq}) may not be hard, as with long exposure experiments the energy profile of signal events might be discernible from its background \cite{Wyenberg:2018eyv,Carew:2023qrj}.  

In a nutshell, if $\chi$ survives accidentally till date, it can be probed in future direct detection experiments with improved sensitivities and long exposure time. 

%%%%%%%%%%%%%%%%%%%%%%%%%%%%%%%%%%%%%%%

\section*{Acknowledgment}
We thank Satyanarayan Mukhopadhyay for various helpful conversations since the outset of this work. We also thank Rohan Pramanick, Sourav Gope and Utpal Chattopadhyay for computational help and Sougata Ganguly, Avirup Ghosh and Tarak Nath Maity for interesting discussions. This work is supported by the Institute Fellowship provided by the Indian Association for the Cultivation of Science (IACS), Kolkata.

%%%%%%%%%%%%%%%%%%%%%%%%%%%%%%%%%%%%%%%%%%%%%%%%%%%%%%%

\appendix
\section{Calculation of loop-induced scattering of neutron (electron) and $\chi$}
\label{app}
\begin{figure}[H]
\centering
\includegraphics[scale=0.6]{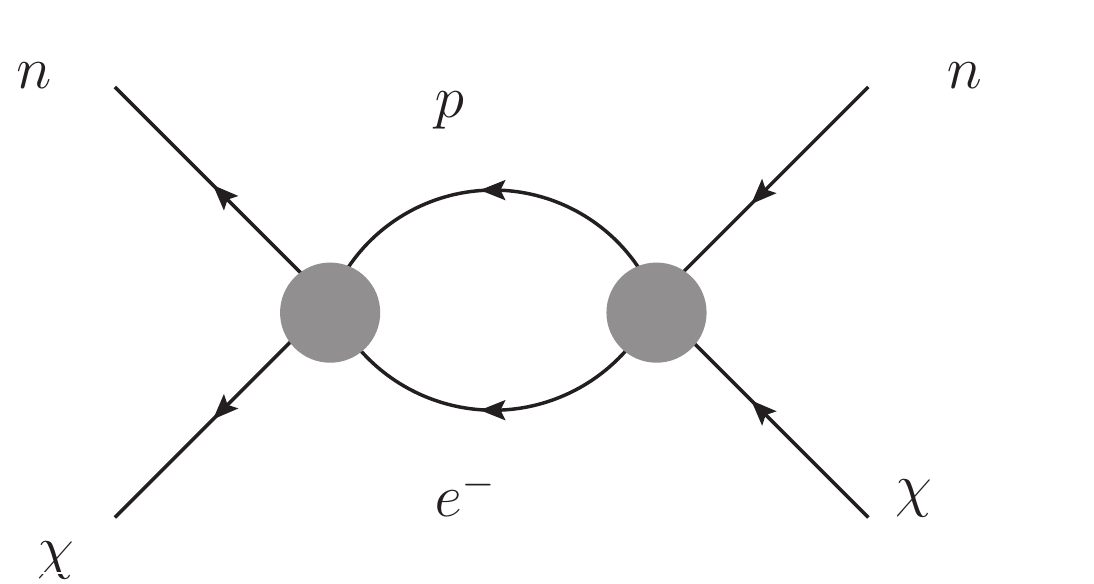}
\includegraphics[scale=0.6]{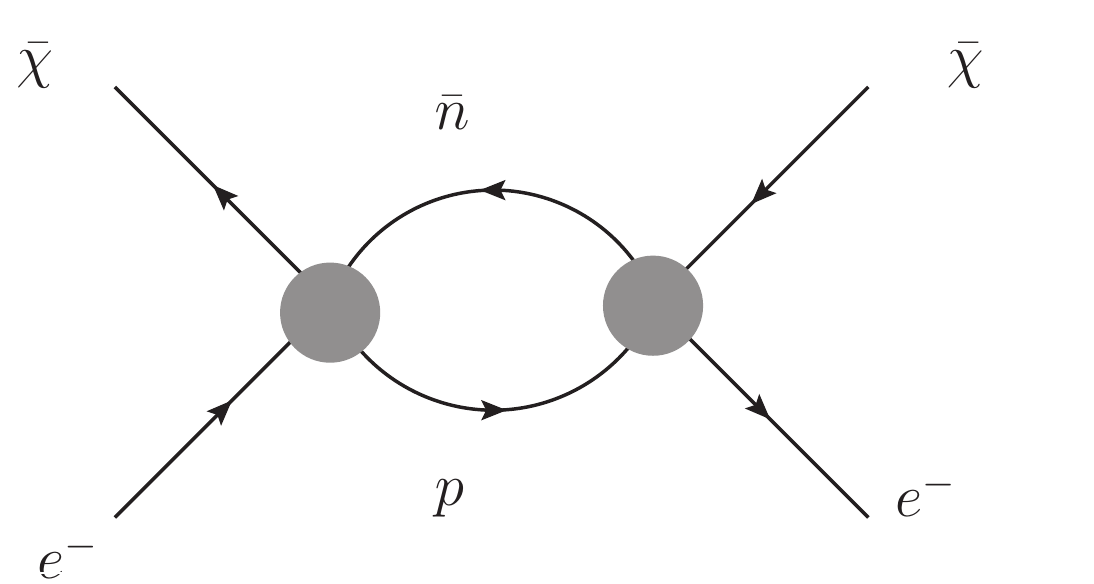}
\caption{Feynman graphs for elastic scattering of neutron (electron) and $\chi$ generated at one-loop level from the co-annihilation operator described in Eq.\ref{eq:model}.}
\label{fig:loop}
\end{figure}

To start with, we had only operator that enables the co-annihilation of $\chi$ and neutron producing a proton and an electron in the final state. This operator itself induces an effective operator at one-loop level, responsible for the elastic scattering of neutron (electron) and $\chi$ as shown in Fig.\ref{fig:loop}, i.e. relevant for direct detection experiments. 
The amplitude for elastic scattering, $n(p_1)+\chi(p_2) \rightarrow n(p_3)+\chi(p_4)$ calculated from Eq.\ref{eq:model} is given by,
\begin{align}
iM = \left(-i \frac{G_D}{\sqrt{2}}\right)^2\int \frac{d^4 k}{(2\pi)^4} \bigg[\bar{u}(3)\,\Gamma^\mu \,\frac{i(\slashed{k}+m_p)}{k^2-m^2_p}\,\Gamma^\nu u(1)\bigg]\, \bigg[\bar{u}(4)\,\Gamma_\mu \,\frac{i(\slashed{P}-\slashed{k}+m_e)}{(P-k)^2-m^2_e}\,\Gamma_\nu u(2)\bigg]
\end{align}
where $u(i)=u(p_i)$ and $P^2=(p_1+p_2)^2=s$, while $p_i$ is the on-shell four momentum. For $\Gamma^a = \gamma^a (I-\gamma_5)$, the form used for all calculations, we get $\Gamma^a \Gamma^b = 0$. Then the amplitude takes rather simpler form, i.e.
\begin{align}
iM= \left(\frac{G_D}{\sqrt{2}}\right)^2\int \frac{d^4 k}{(2\pi)^4} \frac{k_\rho \left(P-k\right)_\sigma}{(k^2-m^2_p)((P-k)^2-m^2_e)}\bigg[\bar{u}(3)\,\Gamma^\mu \,\gamma^\rho\Gamma^\nu u(2)\bigg]\, \bigg[\bar{u}(4)\,\Gamma_\mu \,\gamma^\sigma\Gamma_\nu u(1)\bigg]
\label{eq:loopamp}
\end{align}
The above integral is divergent in general, which is expected in an effective theory with coupling of negative mass dimension. Therefore, we put a cut-off scale, $\Lambda$ and calculate the amplitude as
\begin{align}
iM= \frac{i G^2_D}{32\pi^2}\left(A P_\rho P_\sigma+B\frac{\Lambda^2}{4}\eta_{\rho\sigma}\right) \bigg[\bar{u}(3)\,\Gamma^\mu \,\gamma^\rho\Gamma^\nu u(1)\bigg]\, \bigg[\bar{u}(4)\,\Gamma_\mu \,\gamma^\sigma\Gamma_\nu u(2)\bigg]
\end{align} 
where $A$ and $B$ given by Eq.\ref{eq:loopfactor} are complex numbers in general, to be calculated numerically. 
\begin{align}
A&=\int^1_0 dx\, x(1-x)F_1(x),~~ B=\int^1_0 dx\, F_2 (x)\nonumber\\
F_1(x)&= \log \left(\frac{\Lambda^2+\Delta}{\Delta}\right) - \frac{\Lambda^2}{\Lambda^2+\Delta}\nonumber\\ 
F_2(x)& = 1+ \frac{\Delta}{\Lambda^2+\Delta}-\frac{2\Delta}{\Lambda^2}\log\left(\frac{\Lambda^2+\Delta}{\Delta}\right)\nonumber\\
\Delta &=m^2_e(1-x)+m^2_p \,x - x(1-x)P^2
\label{eq:loopfactor}
\end{align}
Now, this is the most general structure for elastic scattering at hand from our effective theory. In direct detection experiments, the stringent bounds comes from the spin-independent operators. Therefore, we extract the spin-independent part ($M_{SI}$) from the above expression as the following using the relation, $\Gamma^\mu \gamma^\rho \Gamma^\nu = 4\gamma^\mu \eta^{\rho\nu} (I-\gamma_5)- 2\gamma^{\mu}\gamma^\nu \gamma^\rho (I-\gamma_5)$. 

\begin{eqnarray}
iM &=&\frac{i G^2_D}{32\pi^2}\left(A P_\rho P_\sigma+B\frac{\Lambda^2}{4}\eta_{\rho\sigma}\right)\bigg[\bar{u}(3)\bigg(4\gamma^\mu \eta^{\rho\nu} (I-\gamma_5)- 2\gamma^{\mu}\gamma^\nu \gamma^\rho (I-\gamma_5)\bigg)u(1)\bigg] \nonumber\\
&&\hspace{4cm}\bigg[\bar{u}(4)\bigg(4\gamma_\mu \delta^\sigma_\nu (I-\gamma_5)- 2\gamma_{\mu}\gamma_\nu \gamma^\sigma (I-\gamma_5)\bigg)u(2)\bigg]\nonumber\\
iM_{SI} &=& \frac{i G^2_D}{2\pi^2}\left(A P^2 + B \Lambda^2 \right)\bigg[\bar{u}(3)\gamma^\mu u(1)\bigg]\bigg[\bar{u}(4)\gamma_\mu u(2)\bigg]
\label{eq:direct}
\end{eqnarray}
In direct detection experiments, this elastic scattering is considered in the non-relativistic regime, in which the momentum transfer is small compared to particle masses. Therefore, we can approximate Dirac fermion as \cite{Lin:2019uvt}, $u(k)\approx \sqrt{2m}(\xi_s~ \xi_s)^T$, where $\xi_s$ is a spinor with $\sum_s \xi_s \xi^\dagger_s = I$. Consequently, the vector operator in \ref{eq:direct} can be approximated as the scalar one, i.e.
\begin{align}
\bar{u}(p)\gamma^\mu u(p') \rightarrow 2m ~\xi^\dagger_{s'} \xi_s 
\end{align}
Now, setting the cut-off scale to be $\Lambda = (m_n+m_\chi)$, which is also the center-of-mass energy ($\sqrt{P^2}$), we can write the spin-independent effective operator as the following \cite{Lin:2019uvt,Lisanti:2016jxe}. 
\begin{align}
M_{SI} = \frac{G^2_D}{\pi^2}(m_n+m_\chi)^2 \left( 4m_n m_\chi~ \xi^\dagger_{s'}(n) \xi_s(n) \xi^\dagger_{t'}(\chi) \xi_t (\chi) \right),
\end{align}
where $A,B$ is taken as $\mathcal{O}(1)$ numbers. Now, the scattering cross-section in the non-relativistic assumption is given by,
\begin{align}
\sigma_{\chi n}&= \frac{1}{4m_\chi m_n v}\int \frac{d^3 p_3}{(2\pi)^3 2m_n} \frac{d^3 p_2}{(2\pi)^3 2 m_\chi}(2\pi)^4 \delta^4(p_1+p_2-p_3-p_4)\overline{|M_{SI}|^2}\nonumber\\
&=\frac{\overline{|M_{SI}|^2}}{32 \pi m^2_\chi m^2_n v^2}\int^{2\mu_{\chi n}v}_0 p_3 ~dp_3 \int ~d\cos\theta ~\delta \left(\cos\theta -\frac{p_3}{2\mu_{\chi n} v}\right)
\label{eq:directcal}
\end{align}
where the spin-averaged probability of the elastic scattering, $\overline{|M_{SI}|^2} = \frac{G^4_D}{\pi^4}(4 m_\chi m_n (m_\chi+m_n)^2)^2$ and $v$ is the initial velocity of $\chi$ in the rest frame of nucleon. The momentum of the recoiled nucleon in the non-relativistic approximation is given by,$|p_3| = 2\mu_{\chi n} v \cos\theta$, hence comes the upper limit of $p_3$ shown in Eq.\ref{eq:directcal}. $\mu_{\chi n} (=m_\chi m_n/(m_\chi+m_n))$ is the reduced mass of the system   . Finally, the spin-independent $\chi-n$ elastic cross-section ($\sigma_{\chi n}$) turns out to be  
\begin{align}
\sigma_{\chi n} \approx \frac{ G^4_D}{\pi^5} \bigg(m_\chi m_n (m_\chi+m_n) \bigg)^2 .
\end{align}
Now, we can also calculate $\chi-e$ scattering at the one-loop level using same methodology, which produces similar result as in \ref{eq:loopamp}, replacing $m_e$ by $m_n$ in one of the propagators. The loop factors are also modified as $\Delta$ becomes $\tilde{\Delta}=m^2_n (1-x)+m^2_p x - x(1-x)P^2 \approx m^2_n-x(1-x)P^2$. Following similar steps as earlier ( \ref{eq:direct}-\ref{eq:directcal}), we get the spin-independent $\chi-e$ elastic cross-section ($\sigma_{\chi e}$) to be,
\begin{align}
\sigma_{\chi e} \approx \frac{ G^4_D}{\pi^5} \frac{m^2_\chi m^2_e}{(m_\chi+m_e)^2} (m_\chi+m_n)^4 .
\end{align}
To note, we have set same cut-off scale as before, which brings the factor of $(m_\chi+m_n)^4$. The ratio of two scattering cross-sections for $m_\chi>>m_e$ becomes 
\begin{align}
\frac{\sigma_{\chi e}}{\sigma_{\chi n}}\approx \frac{m^2_e}{m^2_\chi m^2_n}  (m_\chi+m_n)^2.
\end{align}

\end{document}